\documentclass[aps,prd,print,groupedaddress,nofootinbib,eqsecnum,onecolumn]{revtex4}
\usepackage{eurosym}
\usepackage{amsmath}
\usepackage{amsthm}
\usepackage{amssymb}
\usepackage{graphicx}
\usepackage{slashed}
\usepackage{hyperref}
\usepackage{epstopdf}

\theoremstyle{plain}

\theoremstyle{definition}

\begin{document}

\author{Michael Rios}
\email{mrios@dyonicatech.com}
\affiliation{Dyonica ICMQG,\\
Los Angeles, CA, USA}
\author{Alessio Marrani}
\email{alessio.marrani@pd.infn.it}
\affiliation{Centro Studi e Ricerche Enrico Fermi, Roma, Italy\\
and \\
Dipartimento di Fisica e Astronomia 'Galileo Galilei', Univ. di Padova, \\
and \\
INFN, Sez. di Padova, Italy}
\author{David Chester}
\email{dchester@ucla.edu}
\affiliation{Department of Physics and Astronomy,\\
UCLA, Los Angeles, CA, USA}
\title{Exceptional Super Yang-Mills in $27+3$ and Worldvolume M-Theory}

\begin{abstract}
Bars and Sezgin have proposed a super Yang-Mills theory in $D=s+t=11+3$
space-time dimensions with an electric 3-brane that generalizes the 2-brane
of M-theory. More recently, the authors found an infinite family of
exceptional super Yang-Mills theories in $D=(8n+3)+3$ via the so-called
Magic Star algebras. A particularly interesting case occurs in signature $%
D=27+3$, where the superalgebra is centrally extended by an electric
11-brane and its 15-brane magnetic dual. The worldvolume symmetry of the
11-brane has signature $D=11+3$ and can reproduce super Yang-Mills theory in
$D=11+3$. Upon reduction to $D=26+2$, the 11-brane reduces to a 10-brane
with $10+2$ worldvolume signature. A single time projection gives a $10+1$
worldvolume signature and can serve as a model for $D=10+1$ M-theory as a
reduction from the $D=26+1$ signature of the bosonic M-theory of Horowitz
and Susskind; this is further confirmed by the reduction of chiral $(1,0)$, $%
D=11+3$ superalgebra to the $\mathcal{N}=1$ superalgebra in $D=10+1$, as
found by Rudychev, Sezgin and Sundell some time ago. Extending previous
results of Dijkgraaf, Verlinde and Verlinde, we also put forward the
realization of spinors as total cohomologies of (the largest spatially
extended) branes which centrally extend the $(1,0)$ superalgebra underlying
the corresponding exceptional super Yang-Mills theory. Moreover, by making
use of an \textquotedblleft anomalous" Dynkin embedding, we strengthen
Ramond and Sati's argument that M-theory has hidden Cayley plane fibers.

$Keywords:$ Super Yang-Mills, Exceptional Periodicity, 11-brane, M-theory.
\end{abstract}

\maketitle

\bigskip  \bigskip

\tableofcontents


\newpage

\setcounter{page}{1}



\section{Introduction}

After Witten's introduction of M-theory \cite{WittenM} in\footnote{%
\textquotedblleft $s$" and \textquotedblleft $t$" denote the number of
spacelike resp. timelike dimensions throughout.} $D=s+t=10+1$ space-time
dimensions, Vafa proposed F-theory in $10+2$ \cite{VafaF}. Bars went
further, generalizing F-theory in $10+2$ and $11+1$ \cite{Hulld} by studying
S-theory in $11+2$ \cite{BarsS,2} and super Yang-Mills (SYM) theory in $11+3$%
, with subsequent further investigations by Sezgin \textit{et al.} \cite%
{1,1bis}. Pushing beyond, Nishino defined SYM's in signature $D=s+t=\left(
9+m\right) +\left( 1+m\right) $, for arbitrary $m\in \mathbb{N}\cup \left\{
0\right\} $ \cite{6}. Using the algebraic structure of \textit{Exceptional
Periodicity} \cite{EP1,EP2,EP3,EP4,EP5}, in \cite{EYM} the authors defined
infinite families of \textit{exceptional} $\left( 1,0\right) $ SYM theories;
among these, a family in $D=s+t=(8n+3)+3$ that directly generalizes Sezgin's
SYM in $11+3$. It is here worth recalling that the structure of SYM\footnote{%
It should be recalled that the Bars-Sezgin SYM theories in $D>10$ are not
Lorentz covariant, and so are the generalizations presented here.} in $11+3$%
, with a $\mathbf{64}$-dimensional Majorana-Weyl (MW) semispinor\footnote{%
It is amusing to observe that the semispinor $\mathbf{64}$ of $Spin(14)$
recently appeared in the $X_{1}$ algebraic structure of the Vogel plane in
\cite{Vogel}.}, interestingly arises in a certain 5-grading of
\textquotedblleft extended Poincar\'{e} type" of $\mathfrak{e}_{8(-24)}$ and
has found use in unification models (see e.g. \cite{Percacci}).

In this work, we ascend to $D=27+3$ space-time dimensions, in which an
electric 11-brane and its 15-brane magnetic dual arise as central extensions
of the $\left( 1,0\right) $ global supersymmetry algebra. In particular, the
11-brane gives rise to a worldvolume theory with $11+3$ signature, thus
providing a \textit{worldvolume embedding} for the chiral SYM in $11+3$ of
Bars and Sezgin \cite{BarsS,2,1}.

Following the 11-brane in the reduction from $27+3\rightarrow
26+2\rightarrow 26+1$ leads to the reduction of the $D=11+3$ worldvolume of
the 11-brane to a $D=10+1$ worldvolume of a 10-brane, suggesting that
M-theory may be a worldvolume theory; this is further confirmed by the fact
that the reduction of the chiral $(1,0)$ superalgebra in $D=11+3$ down to $%
D=10+1$ contains the $\mathcal{N}=1$ superalgebra pertaining to M-theory
\cite{1bis}. This chain of reductions along the worldvolume of the electric
11-brane yields a natural map of the conjectured \textquotedblleft bosonic
M-theory" of Horowitz and Susskind \cite{BMT} in $26+1$ down to M-theory in $%
10+1$. Moreover, in $26+2$ the electric 10-brane has a 14-brane magnetic
dual (both centrally extending the corresponding $\left( 1,0\right) $ global
superalgebra), and this implies, under reduction to $26+1$, that there
exists a \textit{\textquotedblleft dual"} (worldvolume-realized) M-theory in
$D=s+t=13+1$.

\section{$(1,0)$ SYM in $27+3$ and M-theory}

The $(1,0)$ superalgebra in $D=27+3$ space-time dimensions (corresponding to
the level $n=3$ of Exceptional Periodicity \cite{EP1,EP2,EP3,EP4,EP5}) takes
the form \cite{EYM}
\begin{equation}
27+3:~\left\{ Q_{\alpha },Q_{\beta }\right\} =\left( \gamma ^{\mu \nu \rho
}\right) _{\alpha \beta }Z_{\mu \nu \rho }+\left( \gamma ^{\mu _{1}...\mu
_{7}}\right) _{\alpha \beta }Z_{\mu _{1}....\mu _{7}}+\left( \gamma ^{\mu
_{1}...\mu _{11}}\right) _{\alpha \beta }Z_{\mu _{1}....\mu _{11}}+\left(
\gamma ^{\mu _{1}...\mu _{15}}\right) _{\alpha \beta }Z_{\mu _{1}....\mu
_{15}}.  \label{27+3}
\end{equation}%
Namely, the central extensions are given by a 3-brane, a 7-brane, an
electric 11-brane and its dual, a magnetic 15-brane. Note that the magnetic
duals of the 3-brane and 7-brane, \textit{i.e.} the 23-brane resp. 19-brane,
do \textit{not} centrally extend the algebra (\ref{27+3}); however, they can
be found as the largest spatially extended central charges at $n=5$ resp. $%
n=4$ levels of Exceptional Periodicity \cite{EYM}.

In $D=27+3$, the electric 11-brane has a multi-time worldvolume, with
signature $11+3$, which can be used to provide a worldvolume realization for
the $11+3$ SYM of Bars and Sezgin \cite{BarsS,2,1}. In other words, the
multi-time worldvolume of the electric 11-brane in $27+3$ can support a
corresponding $(1,0)$ superalgebra in $11+3$:
\begin{equation}
11+3:~\left\{ Q_{\alpha },Q_{\beta }\right\} =\left( \gamma ^{\mu \nu \rho
}\right) _{\alpha \beta }Z_{\mu \nu \rho }+\left( \gamma ^{\mu _{1}...\mu
_{7}}\right) _{\alpha \beta }Z_{\mu _{1}....\mu _{7}},  \label{jj}
\end{equation}%
whose reduction to lower dimensions contain both the $D=9+1$ type IIB $(2,0)$
chiral superalgebra and the $D=10+1$ ($\mathcal{N}=1$) M-theory
superalgebra, as discussed by Rudychev, Sezgin and Sundell in \cite{1bis}.

Hence, one can consider a reduction $27+3\longrightarrow 26+1$, and focus on
the corresponding reduction $11+3\longrightarrow 10+1$ of the 11-brane
(multi-time) worldvolume down to the (single-time) 10-brane worldvolume (in $%
26+1$); this latter, also in light of the aforementioned reduction of the $%
(1,0)$, $D=27+3$ chiral superalgebra (\ref{jj}) to the $\mathcal{N}=1$
superalgebra in $D=10+1$ \cite{1bis}, can be used to provide a worldvolume
realization of M-theory.

This simple reasoning yields the following consequences:
\begin{itemize}
\item it puts forward the realization of $10+1$ M-theory as a worldvolume
theory of an electric 10-brane in a higher $26+1$ space-time (pertaining to
bosonic M-theory of Horowitz and Susskind \cite{BMT});

\item as such, it provides a map from the bosonic M-theory in $D=26+1$ to
M-theory in $D=10+1$;

\item we observe that bosonic M-theory can be completed to a two-time theory
in $D=26+2$, in which a $(1,0)$ \textit{exceptional} SYM can be defined,
with central extensions given by \cite{EYM}%
\begin{equation}
26+2:~\left\{ Q_{\alpha },Q_{\beta }\right\} =\left( \gamma ^{\mu \nu
}\right) _{\alpha \beta }Z_{\mu \nu }+\left( \gamma ^{\mu _{1}...\mu
_{6}}\right) _{\alpha \beta }Z_{\mu _{1}....\mu _{6}}+\left( \gamma ^{\mu
_{1}...\mu _{10}}\right) _{\alpha \beta }Z_{\mu _{1}....\mu _{10}}+\left(
\gamma ^{\mu _{1}...\mu _{14}}\right) _{\alpha \beta }Z_{\mu _{1}....\mu
_{14}},
\end{equation}%
i.e. by a 2-brane, a 6-brane, and by an electric 10-brane and its dual, a
magnetic 14-brane. This implies that there exists a theory which is \textit{%
dual} to the worldvolume-realized $10+1$ M-theory embedded in $26+2$, namely
the worldvolume-realized $14+2$ theory reduced to $13+1$ of the magnetic
13-brane embedded in $26+1$; we dub such a $13+1$ theory the \textit{%
\textquotedblleft dual M-theory"}, and we leave its study for future work.
\end{itemize}

While the $D=11+3$ superalgebra has been recovered, it is worth clarifying how this sheds light on M-theory,
which contains M2 and M5 branes. The gravity dual of the M2 brane is described by the ABJM theory,
which describes the near-horizon geometry $AdS_4\times S^7$ with SO(8) R-symmetry \cite{Aharony:2008ug}. The gravity
dual of the M5 brane is described by the 6D (2,0) SCFT, which describes the near-horizon geometry $AdS_7\times S^4$ \cite{Claus:1997cq}.
The near-horizon geometries stem from $SO(3,2)\times SO(8)$ and $SO(6,2)\times SO(5)$, respectively, which both
can be broken from $SO(11,2)$, the signature of S-theory. While a generalization of M-theory with $SO(11,3)$ signature hasn't been discussed,
it is generally understood that supergravities as the low-energy limit of string theory comes from a double copy of super-Yang-Mills theories.

From the work above, we are also suggesting that S-theory could be recovered as a worldvolume theory from $SO(27,2)$. This would give
rise to $AdS_{12}\times S^{15}$ near-horizon geometry with $SO(16)$ R-symmetry with a 10-brane source, which ultimately stems from an 11-brane in $D=27+3$. Stacking these D-branes gives a $U(N)$ super-Yang-Mills gauge symmetry, while a generalized ABJM theory would give $U(N)\times U(N)$ symmetry. However, something more exotic than super-Yang-Mills theory would be anticipated in the full worldvolume theory. Higher-derivative terms would be anticipated to give a non-Abelian Dirac-Born-Infeld theory, similar to how the ABJM model has been
extended with higher-derivative terms \cite{Sasaki:2009ij}.

Also, it is worth mentioning that the higher Kaluza-Klein states stem from Kac-Moody and Virasoro extensions of Lie algebras \cite{Hohm:2006ud}. The tower of graviton states comes from a tensor product or double copy of the Virasoro algebra. In particular, M-theory relates to the Kac-Moody algebra $\mathfrak{e}_{11}$ \cite{West:2001as}. It can also be shown that $\mathfrak{e}_{11}$ contains $\mathfrak{so}_{20}\oplus\mathfrak{su}_2\oplus({\bf 512},{\bf2})$, which is precisely $\mathfrak{e}_7^{(2)}$. Therefore, it can be anticipated that EP algebras provide finite-dimensional, non-Lie truncations of infinite-dimensional Kac-Moody algebras, up to a rescaling of some roots in order to give a closed algebra. The $D=27+3$ theory that contains the full spectrum of M-theory as a worldvolume most likely would require either $\mathfrak{e}_{16}$ or a Virasoro and Kac-Moody extension of the EP algebra $e_8^{(3)}$. We leave the investigation of this very interesting issues for future work.

\section{Exceptional Periodicity and Spinors as Brane Cohomologies}

Through the algebraic structure of Exceptional Periodicity, let us consider
the generalization of the split form $\mathfrak{e}_{8(-24)}$ of the largest
finite-dimensional exceptional Lie algebra $\mathfrak{e}_{8}$ provided by $%
\mathfrak{e}_{8(-24)}^{(3)}$, the corresponding, so-called \textit{Magic
Star algebra} at level $n=3$ \cite{EP1,EP2,EP3,EP4,EP5}:
\begin{eqnarray}
\mathfrak{e}_{8(-24)}^{(3)}:= &&\mathfrak{so}_{28,4}\oplus \mathbf{32768}
\label{a-2-3-pre} \\
&=&\mathbf{30}_{-2}\oplus \left( \mathbf{16384}\right) _{-1}^{\prime }\oplus
(\mathfrak{so}_{27,3}\oplus \mathbb{R})_{0}\oplus \mathbf{16384}_{+1}\oplus
\mathbf{30}_{+2},  \label{a-2-3}
\end{eqnarray}%
where $\mathbf{32768=2}^{\mathbf{15}}$ is the MW semispinor in $28+4$, while
$\mathbf{16384=2}^{\mathbf{14}}$ and $\left( \mathbf{16384}\right) ^{\prime }%
\mathbf{=}\left( \mathbf{2}^{\mathbf{14}}\right) ^{\prime }$ denote the MW
spinor and its conjugate in $D=27+3$. $\mathfrak{e}_{8(-24)}^{(3)}$ (\ref%
{a-2-3-pre}) is the $n=3$ element of the countably, Bott-periodized infinite
sequence of generalizations of $\mathfrak{e}_{8(-24)}$. By denoting with $%
\mathbf{2}^{N-1}$ and $\left( \mathbf{2}^{N-1}\right) ^{\prime }$ the chiral
semispinor representations of $\mathfrak{so}_{2N}$, as well as with $\wedge
^{i}\mathbf{N}$ the rank-$i$ antisymmetric ($i$-form) representation of $%
\mathfrak{so}_{N}$, we recall that, as a vector space, the Clifford algebra $%
Cl(N)$ in $N$ dimensions is isomorphic to the Hodge-de Rahm complex in $N$
dimensions :%
\begin{equation}
\dim _{\mathbb{R}}Cl(N)=2^{N}=\dim _{\mathbb{R}}\left( \mathbf{2}%
^{N-1}\oplus \left( \mathbf{2}^{N-1}\right) ^{\prime }\right) =\sum_{i=0}^{N}%
\binom{N}{i}=\dim _{\mathbb{R}}\left( \bigoplus\limits_{i=0}^{N}\wedge ^{i}%
\mathbf{N}\right) .
\end{equation}%
Thus, in the case under consideration, the $2^{15}$-dimensional MW
semispinor $\mathbf{32768}$ of $\mathfrak{so}_{28,4}$, which branches as $%
\mathbf{16384}\oplus \mathbf{16384}^{\prime }$ under $\mathfrak{so}%
_{28,4}\rightarrow \mathfrak{so}_{27,3}$, can be regarded as the total
cohomology of a 15-brane, which in turn can be identified with the maximally
spatially extended central charge of $\mathcal{N}=(1,0)$ SYM (\ref{27+3}) in
$27+3$ space-time dimensions \cite{EYM} :%
\begin{eqnarray}
\underset{\mathfrak{so}_{27,3}\text{~MW~spinor}}{\mathbf{16384}}
&=&\bigoplus\limits_{i=0}^{15}\wedge ^{2i+1}\mathbf{15}=\underset{\mathfrak{%
so}_{15}\text{-cov.~odd~(co)homology~of~the 15-brane}}{\mathbf{15}\oplus
\mathbf{455}\oplus \mathbf{3,003}\oplus \mathbf{6,435}\oplus \mathbf{5,005}%
^{\prime }\oplus \mathbf{1,365}^{\prime }\oplus \mathbf{105}^{\prime }\oplus
\mathbf{1}}; \\
\underset{\mathfrak{so}_{27,3}~\text{MW conjug. spinor}}{\mathbf{16384}%
^{\prime }} &=&\bigoplus\limits_{i=0}^{15}\wedge ^{2i}\mathbf{15}=\underset{%
\mathfrak{so}_{15}\text{-cov.~even~(co)homology~of~the 15-brane}}{\mathbf{1}%
\oplus \mathbf{105}\oplus \mathbf{1,365}\oplus \mathbf{5,005}\oplus \mathbf{%
6,435}^{\prime }\oplus \mathbf{3,003}^{\prime }\oplus \mathbf{455}^{\prime
}\oplus \mathbf{25}^{\prime }}.
\end{eqnarray}

This is nothing but the $n=3$ case of a general fact, namely that the
(chiral) spinor component $\mathbf{2}^{4n+3}$ of the so-called Magic Star
algebra \cite{EP1,EP2,EP3,EP4,EP5} (see also Fig.1)%
\begin{eqnarray}
\mathfrak{e}_{8(-24)}^{(n)} &=&\mathfrak{so}_{8n+4,4}\oplus \mathbf{2}^{4n+3}
\\
&=&\left( \mathbf{8n+6}\right) _{-2}\oplus \left( \mathbf{2}^{4n+2}\right)
_{-1}^{\prime }\oplus (\mathfrak{so}_{8n+3,3}\oplus \mathbb{R})_{0}\oplus
\mathbf{2}_{+1}^{4n+2}\oplus \left( \mathbf{8n+6}\right) _{+2},
\end{eqnarray}%
can be realized as the \textit{total cohomology} of a $\left( 4n+3\right) $%
-brane, which in turn can be identified with the largest spatially extended
central extension of the $\left( 1,0\right) $ supersymmetry algebra in $%
(8n+3)+3$ space-time dimensions \cite{EYM}%
\begin{equation}
\left( 8n+3\right) +3:~\left\{ Q_{\alpha },Q_{\beta }\right\} =\left( \gamma
^{\mu \nu \rho }\right) _{\alpha \beta }Z_{\mu \nu \rho }+\left( \gamma
^{\mu _{1}...\mu _{7}}\right) _{\alpha \beta }Z_{\mu _{1}....\mu
_{7}}+...+\left( \gamma ^{\mu _{1}...\mu _{n+3}}\right) _{\alpha \beta
}Z_{\mu _{1}....\mu _{4n+3}}.
\end{equation}%
Therefore, the spinor generators of the algebra $\mathfrak{e}_{8(-24)}^{(n)}$
are realized, exploiting the central extensions of the $\left( 1,0\right) $
supersymmetry algebra in $(8n+3)+3$, in terms of brane cohomology.

We stress that this realization extends the results found by Dijkgraaf,
Verlinde and Verlinde \cite{DVV} which, in the BPS quantization of the
5-brane, realized the $\mathbf{16}$ components of the central charge as
fluxes through the odd homology cycles on the five-brane itself :%
\begin{eqnarray}
\underset{\mathfrak{so}_{5,5}\text{~MW~spinor}}{\mathbf{16}}
&=&\bigoplus\limits_{i=0}^{5}\wedge ^{2i+1}\mathbf{5}=\underset{\mathfrak{so%
}_{5}\text{-cov.~odd~(co)homology~of~the 5-brane}}{\mathbf{5}\oplus \mathbf{%
10}^{\prime }\oplus \mathbf{1}}; \\
\underset{\mathfrak{so}_{5,5}~\text{MW conjug. spinor}}{\mathbf{16}^{\prime }%
} &=&\bigoplus\limits_{i=0}^{5}\wedge ^{2i}\mathbf{5}=\underset{\mathfrak{so}%
_{5}\text{-cov.~even~(co)homology~of~the 5-brane}}{\mathbf{1}\oplus \mathbf{%
10}\oplus \mathbf{5}^{\prime }}.
\end{eqnarray}

Since the spinor generators are the very ones responsible for the violation
of the Jacobi identity in Magic Star algebras \cite{EP1,EP2,EP3,EP4,EP5},
this is a further hint that the Lie subalgebras of Magic Star algebras yield
purely bosonic sectors. Moreover, it is worth here reminding that the $n=1$
(trivial) level of Exceptional Periodicity boils down to the fact the spinor
component $\mathbf{128}$ of $\mathfrak{e}_{8(-24)}^{(1)}\equiv \mathfrak{e}%
_{8(-24)}$ can be realized as the total cohomology of the 7-brane which
centrally extends the $(1,0)$ superalgebra in $11+3$ \cite{BarsS,1,1bis,2}.

\begin{figure}[tbp]
\centering
\includegraphics[width=0.5\textwidth]{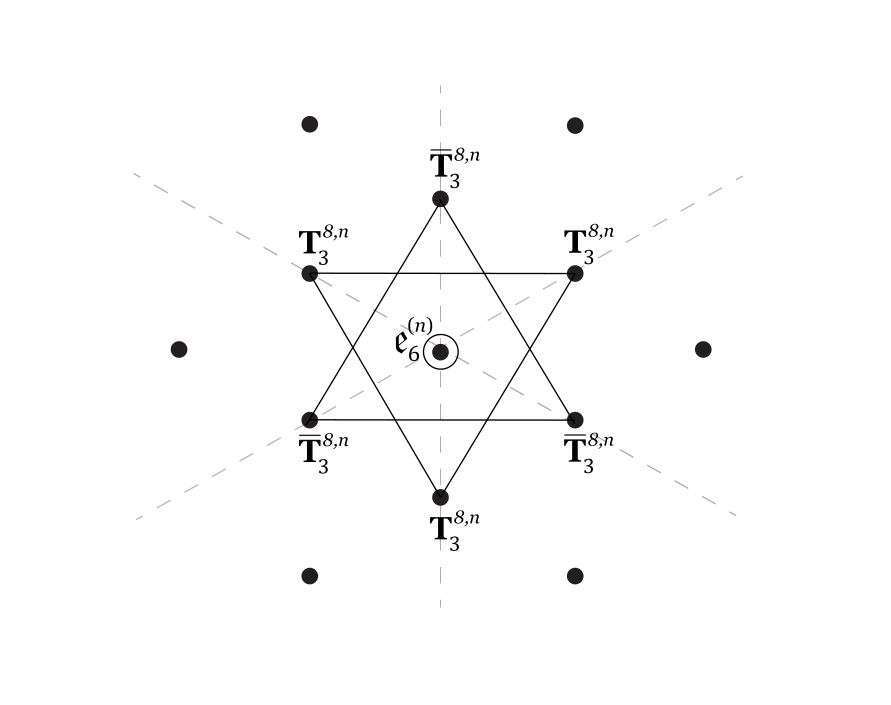}
\caption{The \textit{\textquotedblleft Magic Star"} of Exceptional
Periodicity \protect\cite{EP1} allows for $\mathfrak{e}_{8}^{(n)}$ to be
found as a finite-dimensional, Jacobi-violating generalization of $\mathfrak{%
e}_{8}$ \protect\cite{EP1,EP2,EP3,EP4,EP5}. $\mathbf{T}_{3}^{8,n}$ denotes
the Hermitian part of a T-algebra of rank-3 and of \textit{special type}
\protect\cite{Vinberg}, parametrized by the octonions $\mathbb{O}$ with $%
n\in \mathbb{N}$ \protect\cite{Vinberg,EP5}.}
\label{fig:MagicStarT}
\end{figure}

\section{Brane Actions and Hidden $\mathbb{OP}^{2}$ Fibers in M-Theory}

As resulting from the star-shaped algebraic structure of the Magic Star
algebras \cite{EP1,EP2,EP3,EP4,EP5} (see also Fig.1), we note that the
Hermitian part of the cubic Vinberg's T-algebra\footnote{%
This has been named \textit{HT-algebra} in \cite{EP5}.} \cite{Vinberg} at EP
level $n=3$ (i.e., $\mathfrak{e}_{8(-24)}^{(3)}$) can be Peirce-decomposed
(in a manifestly $\mathfrak{so}_{25,1}$-covariant way) as%
\begin{equation}
\mathbf{T}_{3}^{8,3}=\mathbf{26}\oplus \mathbf{4096}\oplus \mathbf{1},
\label{T}
\end{equation}%
where $\mathbf{26}$ and $\mathbf{4096=2}^{12}$ respectively are the vector
and MW semispinor irreprs. of $\mathfrak{so}_{25,1}$. Adopting Sezgin et
al.'s approach \cite{1,1bis}, the larger symmetry $\mathfrak{so}_{28,4}$ can
be considered as a multi-particle symmetry, for four particles, where
putting all particles but one on-shell yields constant momenta that appear
as null vectors \cite{1bis}; a single particle (described by the algebra $%
\mathbf{T}_{3}^{8,3}$) enjoys $\mathfrak{so}_{25,1}$ symmetry, while two and
three particles acquire enhanced $\mathfrak{so}_{26,2}$ and $\mathfrak{so}%
_{27,3}$ symmetry, respectively. In such a perspective, bosonic M-theory in $%
D=26+1$ can be considered as a single time projection of a two-particle
system with reduced $\mathfrak{so}_{26,2}\rightarrow \mathfrak{so}_{26,1}$
symmetry and Horowitz and Susskind's low energy $D=26+1$ action \cite{BMT}

\begin{equation}
S=\int{d^{27}x\sqrt{-\hat{g}}\left[R(\hat{g})-\frac{1}{48}%
F_{\mu\nu\rho\sigma}F^{\mu\nu\rho\sigma}\right]}
\end{equation}%
captures a 2-brane in this background where $F=dC$, and this reduces to the
bosonic string action in $D=25+1$. As the $D=26+2$ superalgebra admits a
10-brane as well as a 2-brane, one can alternatively consider a $D=26+1$ low
energy action in terms of a 12-form field strength

\begin{equation}
S=\int{d^{27}x\sqrt{-\hat{g}}\left[R(\hat{g})-\frac{1}{2(12!)}%
F_{\mu_1\cdots\mu_{12}}F^{\mu_1\cdots\mu_{12}}\right]}
\end{equation}%
suggesting a (10,1) signature worldvolume M-theory with sixteen transverse
directions. A 2-brane can be immersed in the (10,1) worldvolume as an
M2-brane or in the (26,1) signature bulk, reducing to either a type IIA
string \cite{Duff87} or $D=25+1$ bosonic string \cite{BMT} by
compactification, respectively. The full worldvolume symmetry in $D=26+2$
for the 10-brane is $SO(10,2)$, the signature of F-theory \cite{VafaF}
\footnote{%
This might seem in contrast with the claim that F-theory has signature $11+1$%
, made at the start of the paper. However, the signature of the two
additional dimensions of F-theory with respect to string theory is somewhat
ambiguous due to their infinitesimal character. For example, the
supersymmetry of F-theory on a flat background corresponds to type IIB (i.e.
$(2,0)$) supersymmetry with 32 real supercharges which may be interpreted as
the dimensional reduction of the chiral real 12-dimensional supersymmetry if
its signature is $10+2$. On the other hand, the signature $11+1$ is needed
for the Euclidean interpretation of the compactification spaces (e.g. the
four-folds), and the latter interpretation prevailed in recent years.};
intriguingly, in such a signature the null reduction of a (2,2) brane in $%
D=10+2$ yields to type IIB string theory in $D=9+1$ \cite{Duff88}.

The $D=27+3$ superalgebra permits an 11-brane with 13-form field strength,
which one can use for worldvolume reduction to $D=11+3$, whose $\left(
1,0\right) $ superalgebra yields the $\mathcal{N}=1$ superalgebra in $D=10+1$
as well as the IIA, IIB and heterotic superalgebras in $D=9+1$ \cite{1bis},
with sixteen dimensions transverse to the 11-brane. Moreover, reducing $11+3$
to $11+1$ allows an identification of the 3-brane with the type IIB
D3-brane, as noted by Tseytlin \cite{Tseyt}.

A non-compact real form of a theorem by Dynkin \cite{Dynkin} yields the
maximal and non-symmetric \textquotedblleft anomalous"\footnote{%
This naming goes back to Ramond \cite{ramond3}.} embedding \cite{ramond3}%
\begin{eqnarray}
\mathfrak{f}_{4(-20)} &\subset &\mathfrak{so}_{25,1}  \notag \\
\mathbf{26} &=&\mathbf{26,}
\end{eqnarray}%
under which the vector representation of $\mathfrak{so}_{25,1}$ stays
irreducible, providing the fundamental representation of the minimally
non-compact real form $\mathfrak{f}_{4(-20)}$ of $\mathfrak{f}_{4}$. Since $%
\mathfrak{f}_{4(-20)}$ is the Lie algebra of the derivations of the
Lorentzian version of the exceptional cubic Jordan algebra over the
octonions $J_{1,2}^{\mathbb{O}}$ \cite{GZ-5, Squaring-Magic},%
\begin{equation}
\mathfrak{f}_{4(-20)}=\mathfrak{der}\left( J_{1,2}^{\mathbb{O}}\right) ,
\end{equation}%
it follows that the vector $\mathbf{26}$ of $\mathfrak{so}_{25,1}$ may be
regarded as the traceless part of $J_{1,2}^{\mathbb{O}}$ :%
\begin{equation}
\mathbf{26}\simeq \left( J_{1,2}^{\mathbb{O}}\right) _{0}.
\end{equation}%
The threefold Pierce decomposition of $\left( J_{1,2}^{\mathbb{O}}\right)
_{0}$, corresponding to the branching%
\begin{eqnarray}
\mathfrak{so}_{9} &\subset &\mathfrak{f}_{4(-20)}  \notag \\
\mathbf{1}\oplus \mathbf{9}\oplus \mathbf{16} &\mathbf{=26,}&
\end{eqnarray}%
is geometrically realized as three transverse Hopf maps $S^{7}%
\hookrightarrow S^{15}\rightarrow S^{8}$, and it yields to three cosets of
type%
\begin{equation}
\frac{F_{4(-20)}}{SO(9)},
\end{equation}%
providing the charts of the (non-compact, Riemannian real form of the)
Cayley plane $\mathbb{OP}^{2}$. This, in the worldvolume picture, confirms
and strengthens Ramond and Sati's argument that $D=10+1$ M-theory has hidden
Cayley plane fibers \cite{ramond,ramond2,sati,sati2}. Moreover, the $10+1$
worldvolume realization of M-theory proposed in this paper provides a
natural realization of M-theory on a manifold with boundary, as it was
considered by Horava and Witten in \cite{HW}, in which the cancellation of
anomalies selects the $E_8\times E_8$ gauge symmetry of heterotic string
theory.

Horowitz and Susskind noticed that $D=26+1$ M-theory predicts the existence
of a $2+1$ CFT with $SO(24)$ symmetry \cite{BMT}. Here, we confine ourselves
to noting that the two-particle symmetry $\mathfrak{so}_{26,2}$ is the Lie
algebra of the group $SO(26,2)$, whose decomposition $SO(26,2)\rightarrow
SO(2,2)\times SO(24)$ suggests the existence of an $AdS_{3}\times S^{23}$
geometry, which in turn supports the existence of a $2+1$ CFT with $SO(24)$
symmetry. Along this line, the result that the fourth integral cohomology of
Conway's group $Co_{0}$ \cite{wilsonMonster} is a cyclic group of order 24
\cite{maths!}, as well as the fact that $Co_{0}$ is a maximal finite
subgroup of $SO(24)$ \cite{maths!2}, seem to suggest that such a CFT could
be related to Witten's monster CFT construction for 3D gravity \cite%
{WittenMonster}; in the multi-particle picture, the twenty-four transverse
directions would be discretized as the Leech lattice \cite%
{wilsonLeech,riosLeech}.

\section{\label{Conclusions}Conclusion}

Using the \textit{exceptional} SYM theory in $27+3$ space-time dimensions,
whose $(1,0)$ non-standard global superalgebra can be centrally extended by
an electric 11-brane and its 15-brane magnetic dual \cite{EYM}, we
considered the (multi-time) worldvolume theory of the 11-brane itself as
support for the $(1,0)$ SYM theory in $11+3$ space-time dimensions as
introduced by Bars and Sezgin some time ago \cite{BarsS,2,1}.

As the $(1,0)$ superalgebra in $11+3$ dimensions reduces to the $10+1$ $%
\mathcal{N}=1$ superalgebra, as well as the type IIA, IIB and heterotic
superalgebras in $9+1$ \cite{1bis}, we proposed the reduced (single-time)
10-brane worldvolume theory in $10+1$ as a \textit{worldvolume realization}
of M-theory (this also entails the existence of a would-be \textquotedblleft
dual worldvolume M-theory" realized as a worldvolume theory in $13+1$). In
this framework, the space-time reduction $27+3\longrightarrow 26+1$ yields a
natural map from the conjectured bosonic M-theory of Horowitz and Susskind
\cite{BMT} in $D=26+1$ to $D=10+1$ M-theory. The worldvolume picture is
essential in geometrically explaining the origin of the $E_8\times E_8$
heterotic string; the Horava-Witten domain wall for heterotic M-theory
requires a manifold with boundary in eleven dimensions, which occurs
naturally if the eleven dimensional manifold is itself a brane worldvolume
with boundary.

Moreover, extending the results of \cite{DVV}, we have put forward the
intriguing brane-cohomological interpretation of spinors, and in particular
of the spinor generators of the recently discovered class of Magic Star
algebras \cite{EP4,EP5}, thus entangling the algebraic structure of
Exceptional Periodicity \cite{EP1,EP2,EP3,EP4,EP5} with the central
extensions of exceptional super Yang Mills theories in higher dimensional
space-times.

\textit{Last but not least}, by recalling an \textquotedblleft anomalous"
Dynkin embedding \cite{Dynkin,ramond3}, we identified the vector irrepr. in
twenty-six Lorentzian dimensions as the traceless part of (the real,
Lorentzian version of) the exceptional cubic Jordan algebra over the
octonions, whose Peirce decomposition strengthens Ramond and Sati's argument
that $D=10+1$ M-theory has hidden Cayley plane fibers \cite%
{ramond,ramond2,sati,sati2}.


\begin{thebibliography}{99}
\bibitem{WittenM} E. Witten, \textit{String Theory Dynamics In Various
Dimensions}, Nucl. Phys. \textbf{B443}, 85 (1995), \texttt{\href{https://arxiv.org/abs/hep-th/9503124}%
{arXiv:hep-th/9503124}}.

\bibitem{VafaF} C. Vafa, \textit{Evidence for F-Theory}, Nucl. Phys. \textbf{%
B469}, 403 (1996), \href{https://arxiv.org/abs/hep-th/9602022}{\texttt{%
arXiv:hep-th/9602022}}.

\bibitem{Tseyt} A. A. Tseytlin, \textit{Self-duality of Born-Infeld action
and Dirichlet 3-brane of type IIB superstring theory}, Nucl. Phys. \textbf{%
B469}, 51 (1996), \href{https://arxiv.org/abs/hep-th/9602064}{\texttt{%
arXiv:hep-th/9602064}}.

\bibitem{Duff88} M. P. Blencowe, M. J. Duff, \textit{Supermembranes and the
signature of spacetime}, Nucl. Phys. \textbf{B310}, 387 (1988).

\bibitem{Duff87} M. J. Duff, P. S. Howe, T. Inami, K. S. Stelle, \textit{%
Superstrings in $D=10$ from supermembranes in $D=11$}, Phys. Lett. \textbf{%
B191}, 70 (1987).

\bibitem{Hulld} C. M. Hull, \textit{Duality and the Signature of Space-Time}%
, JHEP \textbf{9811}, 017 (1998), \href{https://arxiv.org/abs/hep-th/9807127}%
{\texttt{arXiv:hep-th/9807127}}.

\bibitem{BarsS} I. Bars, \textit{S-theory}, Phys. Rev. \textbf{D55}, 2373
(1997), \href{https://arxiv.org/abs/hep-th/9607112}{\texttt{%
arXiv:hep-th/9607112}}.

\bibitem{2} I. Bars, \textit{A case for }$\mathit{14}$\textit{\ dimensions},
Phys. Lett. \textbf{B403}, 257 (1997), \texttt{\href{https://arxiv.org/abs/hep-th/9704054}%
{arXiv:hep-th/9704054}}.

\bibitem{1} E. Sezgin, \textit{Super Yang-Mills in }$\mathit{(11,3)}$\textit{%
\ Dimensions}, Phys. Lett. \textbf{B403}, 265 (1997), \texttt{\href{https://arxiv.org/abs/hep-th/9703123}%
{arXiv:hep-th/9703123}}.

\bibitem{1bis} I. Rudychev, E. Sezgin, P. Sundell, \textit{Supersymmetry in
dimensions beyond eleven}, Nucl. Phys. Proc. Suppl. \textbf{68}, 285 (1998),
\texttt{\href{https://arxiv.org/abs/hep-th/9711127}{arXiv:\texttt{%
hep-th/9711127}}}.

\bibitem{6} H. Nishino, \textit{Supersymmetric Yang-Mills Theories in }$%
\mathit{D}\geqslant \mathit{12}$, Nucl. Phys. \textbf{B523}, 450 (1998),
\href{https://arxiv.org/abs/hep-th/9708064}{\texttt{arXiv:hep-th/9708064}}.

\bibitem{EP1} P. Truini, M. Rios, A. Marrani, \textit{The Magic Star of
Exceptional Periodicity}, in : \textit{\textquotedblleft Nonassociative
Mathematics and its Applications"}, Contemporary Mathematics, Vol. \textbf{%
721}, (2019), 277-297, AMS, \texttt{\href{https://arxiv.org/abs/1711.07881}{%
arXiv:1711.07881 [hep-th]}}.

\bibitem{EP2} P. Truini, A. Marrani, M. Rios, \textit{Magic Star and
Exceptional Periodicity: an approach to Quantum Gravity}, J. Phys. Conf.
Ser. \textbf{1194} (2019) no.1, 012106, \href{https://arxiv.org/abs/1811.11202}%
{\texttt{arXiv:1811.11202 [hep-th]}}.

\bibitem{EP3} A. Marrani, P. Truini, M. Rios, \textit{The Magic of Being
Exceptional}, J. Phys. Conf. Ser. \textbf{1194} (2019) no.1, 012075, \href{https://arxiv.org/abs/hep-th/9708064}%
{\texttt{arXiv:1811.11208 [hep-th]}}.

\bibitem{EP4} P. Truini, A. Marrani, M. Rios, \textit{Exceptional
Periodicity and Magic Star Algebras. I : Foundations}, \href{https://arxiv.org/abs/1909.00357}%
{\texttt{arXiv:1909.00357 [math.RT]}}.

\bibitem{EP5} P. Truini, A. Marrani, M. Rios, \textit{Exceptional
Periodicity and Magic Star Algebras. II : Gradings and HT-Algebras}, \href{https://arxiv.org/abs/1910.07914}%
{\texttt{arXiv:1910.07914 [math.RT]}}.

\bibitem{EYM} M. Rios, A. Marrani, D. Chester, \textit{The Geometry of
Exceptional Super Yang-Mills Theories}, \ Phys. Rev. \textbf{D99} (2019)
046004, \texttt{\href{https://arxiv.org/abs/1811.06101}{arXiv:1811.06101
[hep-th]}}.

\bibitem{Vogel} R. L. Mkrtchyan, \textit{On the map of Vogel's plane}, Lett.
Math. Phys. \textbf{106} (2016) no.1, 57-79, \texttt{\href{https://arxiv.org/abs/1209.5709}%
{arXiv:1209.5709 [math-ph]}}.

\bibitem{Percacci} R. Percacci, \textit{Gravity from a Particle Physicists'
perspective}, PoS ISFTG \textbf{2009}, 011 (2009), \href{https://arxiv.org/abs/0910.5167}%
{\texttt{arXiv:0910.5167 [hep-th]}}.

\bibitem{BMT} G.T. Horowitz, L. Susskind, \textit{Bosonic M-theory}, J.
Math. Phys. \textbf{42}, 3152 (2001), \texttt{\href{https://arxiv.org/abs/hep-th/0012037}%
{arXiv:hep-th/0012037}}.

\bibitem{Aharony:2008ug}
O.~Aharony, O.~Bergman, D.~L.~Jafferis and J.~Maldacena, \textit{$\mathcal{N}=6$ superconformal Chern-Simons-matter theories, M2-branes and their gravity duals}, JHEP \textbf{10}, 091 (2008), \texttt{\href{https://arxiv.org/abs/0806.1218}%
{arXiv:0806.1218 [hep-th]}}.

\bibitem{Claus:1997cq}
P.~Claus, R.~Kallosh and A.~Van Proeyen, \textit{M five-brane and superconformal (0,2) tensor multiplet in six-dimensions}, Nucl. Phys. \textbf{B518}, 117-150 (1998), \texttt{\href{https://arxiv.org/abs/hep-th/9711161}%
{arXiv:hep-th/9711161 [hep-th]}}.

\bibitem{Sasaki:2009ij}
S.~Sasaki, \textit{On Non-linear Action for Gauged M2-brane}, JHEP \textbf{02}, 039 (2010), \texttt{\href{https://arxiv.org/abs/0912.0903}%
{arXiv:0912.0903 [hep-th]}}.

\bibitem{Hohm:2006ud}
O.~Hohm, \textit{Gauged diffeomorphisms and hidden symmetries in Kaluza-Klein theories}, Class. Quant. Grav. \textbf{24}, 2825-2844 (2007), \texttt{\href{https://arxiv.org/abs/hep-th/0611347v3}%
{arXiv:hep-th/0611347 [hep-th]}}.

\bibitem{West:2001as}
P.~C.~West, \textit{$E_{11}$ and M theory}, Class. Quant. Grav. \textbf{18}, 4443-4460 (2001),
\texttt{\href{https://arxiv.org/abs/hep-th/0104081}%
{arXiv:hep-th/0104081 [hep-th]}}.


\bibitem{DVV} R. Dijkgraaf, E. P. Verlinde and H. L. Verlinde, \textit{BPS
quantization of the five-brane}, Nucl. Phys. \textbf{B486}, 89 (1997), \href{https://arxiv.org/abs/hep-th/9604055}%
{\texttt{arXiv:hep-th/9604055}}.

\bibitem{Vinberg} E.B. Vinberg, \textit{The theory of Convex Homogeneous
Cones}, in : Transactions of the Moscow Mathematical Society for the year
1963, 340-403, American Mathematical Society, Providence RI 1965.

\bibitem{Dynkin} E. Dynkin, \textit{Maximal Subgroups of the Classical Groups%
}, Tr. Mosk. Mat. Obshchestva. \textbf{1}, 39 (1952) [English transl. :
Amer. Math. Soc. Transl. Ser. \textbf{2} 6, 245 (1957)].

\bibitem{ramond3} P. Ramond, \textit{Exceptional groups and physics},
plenary talk at the \textit{24th International Colloquium on Group
Theoretical Methods in Physics (GROUP 24)}, 15-20 Jul 2002, Paris, France;
published in Inst. Phys. Conf. Ser. \textbf{173} (2003), \href{https://arxiv.org/abs/hep-th/0301050}%
{\texttt{arXiv:hep-th/0301050}}.

\bibitem{GZ-5} M. G\"{u}naydin, M. Zagermann, \textit{Unified
Maxwell-Einstein and Yang-Mills-Einstein supergravity theories in
five-dimensions}, JHEP \textbf{0307} (2003) 023, \href{https://arxiv.org/abs/hep-th/0304109}%
{\texttt{arXiv:hep-th/0304109}}.

\bibitem{Squaring-Magic} S. L. Cacciatori, B. L. Cerchiai, A. Marrani,
\textit{Squaring the Magic}, Adv. Theor. Math. Phys. \textbf{19} (2015)
923-954, \href{https://arxiv.org/abs/1208.6153}{\texttt{arXiv:1208.6153
[math-ph]}}.

\bibitem{ramond} T. Pengpan, P. Ramond, \textit{M(ysterious) Patterns in
SO(9)}, Phys. Rept. \textbf{315} (1999) 137-152, \texttt{\href{https://arxiv.org/abs/hep-th/9808190}%
{arXiv:hep-th/9808190}}.

\bibitem{ramond2} P. Ramond, \textit{Boson-Fermion Confusion: The String
Path To Supersymmetry}, Nucl. Phys. Proc. Suppl. \textbf{101} (2001) 45-53,
\texttt{\href{https://arxiv.org/abs/hep-th/0102012}{arXiv:hep-th/0102012}}.

\bibitem{sati} H. Sati, \textit{On the geometry of the supermultiplet in
M-theory}, Int. J. Geom. Meth. Mod. Phys. \textbf{8} (2011) 1-33, \texttt{\
\href{https://arxiv.org/abs/0909.4737}{arXiv:0909.4737 [hep-th]}}.

\bibitem{sati2} H. Sati, \textit{$\mathbb{OP}^{2}$ bundles in M-theory},
Commun. Num. Theor. Phys. \textbf{3} (2009) 495-530, \texttt{\href{https://arxiv.org/abs/0807.4899}%
{arXiv:0807.4899 [hep-th]}}.

\bibitem{HW} P. Horava, E. Witten, \textit{Eleven-dimensional supergravity
on a manifold with boundary}, Nucl. Phys. \textbf{B475} (1996) 94-114,
\texttt{\href{https://arxiv.org/abs/hep-th/9603142}{arXiv:hep-th/9603142}}.

\bibitem{wilsonMonster} P. E. Holmes, R. A. Wilson, \textit{A new computer
construction of the Monster using 2-local subgroups}, J. Lond. Math. Soc.
\textbf{67} (2003) no. 2, 349-364.

\bibitem{maths!} T. Johnson-Freyd, D. Treumann, $H^{4}(Co_{0};\mathbf{Z})$ $=%
\mathbf{Z}/24$, Int. Math. Res. Not. rny\textbf{219}, \href{https://arxiv.org/abs/1707.07587}%
{\texttt{arXiv:1707.07587 [math.GR]}}.

\bibitem{maths!2} J. F. Duncan, \textit{Super-moonshine for Conway's largest
sporadic group}, Duke Math. J. \textbf{139 }(2) 255-315 (2007), \href{https://arxiv.org/abs/math/0502267}%
{\texttt{arXiv:math/0502267}}.

\bibitem{WittenMonster} E. Witten, \textit{Three-Dimensional Gravity
Revisited}, \href{https://arxiv.org/abs/0706.3359}{\texttt{arXiv:0706.3359
[hep-th]}}.

\bibitem{wilsonLeech} R. A. Wilson, \textit{Octonions and the Leech lattice}%
, J. Algebra \textbf{322} (2009) 2186-2190.

\bibitem{riosLeech} M. Rios, \textit{U-Duality and the Leech Lattice},
\texttt{\href{https://arxiv.org/abs/1307.1554}{arXiv:1307.1554 [hep-th]}}.
\end{thebibliography}
\end{document}